\documentclass{PoS}
\usepackage{setspace}
\usepackage{slashed}
\usepackage{verbatim}    
\usepackage{graphicx} 
\usepackage{wrapfig}

\usepackage{epsf}
\usepackage{epsfig}
\usepackage{epstopdf}
\usepackage{amsmath}
\usepackage{amsfonts,amssymb}
\usepackage{cleveref}

\newcommand{\be}{\begin{equation}}
\newcommand{\ee}{\end{equation}}
\newcommand{\bea}{\begin{eqnarray}} 
\newcommand{\eea}{\end{eqnarray}}
\newcommand{\MSbar}{{\overline{\rm MS}}}

\newcommand{\la}{\lambda}
\newcommand{\Dslash}{{\not{\hspace{-0.1cm}D}}}

\title{Perturbative study of the Gluino-Glue operator in SYM}
\ShortTitle{Gluino-Glue operator in SYM}

\author{\speaker{M.~Costa}$^{, \,a}$, G.~Panagopoulos$^b$, H.~Panagopoulos$^a$, G.~Spanoudes$^{a,\, c}$, H.~Herodotou$^a$, P.~Philippides$^a$\\
	\llap{}$^a$Department of Physics, University of Cyprus, Nicosia, CY-1678, Cyprus\\
	$^b$Department of Physics, Stanford University, CA 94305-2004, USA\\
	$^c$Present address: Computation-based Science and Technology Research
	Center, The Cyprus Institute, 20 Kavafi Str., Nicosia
	2121, Cyprus.\\
	{\rm E-mail}: \email{kosta.marios@ucy.ac.cy}, \email{gpanago@stanford.edu}, \email{panagopoulos.haris@ucy.ac.cy}, \email{g.spanoudis@cyi.ac.cy},
	\email{herodotos.herodotou@ucy.ac.cy}, \email{phivos.philippides@ucy.ac.cy}}

\abstract{We investigate the renormalization of the Gluino-Glue operator, using both Lattice Perturbation Theory (LPT) and a Gauge Invariant Renormalization Scheme (GIRS). The latter scheme involves gauge-invariant Green's functions of two operators at different space-time points, which can be also computed via numerical simulations. There is no need to fix a gauge and the mixing with gauge noninvariant operators is inconsequential. We calculate perturbatively the conversion factor relating GIRS with the Modified Minimal Subtraction scheme. On the other hand, the Gluino-Glue operator being mixes with several gauge noninvariant operators which have the same quantum numbers. The determination of the mixing matrix on the lattice demands the calculation of 2-pt and 3-pt Green's functions with external gluon, gluino and ghost fields using LPT. We compute at one-loop order the renormalization of the Gluino-Glue operator and all operator mixing coefficients.}

\FullConference{%
	The 38th International Symposium on Lattice Field Theory, LATTICE2021
	26th-30th July, 2021
	Zoom/Gather@Massachusetts Institute of Technology
}

\begin{document}
	\maketitle
	
	\section{Introduction}
	
	Supersymmetric strongly coupled theories have influenced experimental particle physics and evidence of SUSY as well as its predictions are still under investigation. SUSY is expected to emerge at very high energies and it provides dark matter candidates, arising from the lightest supersymmetric particles. In addition to that, supersymmetric extensions of the Standard Model would resolve the hierarchy problem. In this work, we considered the ${\cal N}=1$ supersymmetric Yang-Mills (SYM) theory, which describes the strong interactions between gluons ($u_\mu$) and gluinos ($\la$), the superpartners of the gluons. SYM shares some of the fundamental properties of supersymmetric theories containing quarks and squarks, while at the same time it is amenable to high-accuracy non-perturbative investigations; it is thus an ideal forerunner to the future study of theories containing more superfields.
	
	A fundamental ingredient in these investigations is the ``Gluino-Glue'' composite operator, ${\cal O}_{Gg}$. In the present work we study thoroughly the renormalization and mixing of this operator, to the first perturbative order. In particular, we present two separate calculations: 
	1) We compute, in dimensional regularization, the conversion factors relating the $\MSbar$ scheme to an intermediate gauge-invariant coordinate-space scheme. In this second scheme, there is no mixing with gauge variant operators and no need to fix the gauge. Further, all necessary correlation functions can be computed nonperturbatively in simulations, again without need for gauge fixing or ghost fields. However, a downside of this scheme is that the calculations to order $g^{2n}$ requires evaluation of $(n+1)$-loop Feynman diagrams.
	2) We use lattice perturbation theory and compute, to one loop, various 2- and 3-point functions of ${\cal O}_{Gg}$. We consider mixing with all relevant gauge-noninvariant operators, which contain also ghost fields.
	
	For a comprehensive presentation of our results, along with detailed explanations and a longer list of references, we refer to our publications~\cite{Costa:2021pfu, Costa:2020keq}.
	
	\section{Gluino-Glue Operator}
	In studying the properties of light bound states, the main observables in SYM would be particles made up of gluons and gluinos. An important object in such a study is the Gluino-Glue operator, defined as follows:
	\be
	{\cal O}_{Gg} = \sigma_{\mu \nu} \,{\rm{tr}}_c (\, u_{\mu \nu} \lambda ), \quad \sigma_{\mu \nu}=\frac{1}{2} [\gamma_{\mu},\gamma_{\nu}],\quad  u_{\mu \nu} = \partial_{\mu}u_{\nu}-\partial_{\nu}u_{\mu}+ i g [u_{\mu},u_{\nu}]
	\label{GgO}
	\ee
	This operator has the lowest possible dimensionality (7/2) compatible with gauge invariance and being at the same time fermionic.  Acting on the vacuum, ${\cal O}_{Gg}$ is expected to excite a light bound state of the theory, which is a potential supersymmetric partner of the gluinoballs (${\rm{tr}}_c(\bar \la \la)$, ${\rm{tr}}_c(\bar \la \gamma_5 \la)$, $\cdots)$ and the glueballs (${\rm{tr}}_c(u_{\mu \nu}u_{\mu \nu})$, ${\rm{tr}}_c(u_{\mu \nu} \tilde u_{\mu \nu})$, $\cdots$)~\cite{Veneziano:1982ah}. The Gluino-Glue operator, being composite, could in principle mix with four classes of operators having the same quantum numbers. The four classes are as follows: {\bf Class G} are gauge invariant operators. {\bf Class A} are operators which are not gauge invariant but are the  BRST variation of some other operators: ${\cal O}_{A}$. {\bf Class B} operators vanish by the equations of motion: ${\cal O}_{B}$. {\bf Class C} are operators which are not linear combinations of class G, A and B: ${\cal O}_{C}$. Below, we present all candidate operators which can mix with ${\cal O}_{Gg}$ ($\alpha$: gauge fixing parameter, $c$: ghost field):
	\bea
	{\cal O}_{A1} &=& \frac{1}{\alpha} {\rm{tr}}_c (\la\,  \partial_\mu u_\mu) - ig\, {\rm{tr}}_c(\la [c, \bar{c}]) \nonumber  \\[0,1ex] \nonumber 
	{\cal O}_{B1} &=&  {\rm{tr}}_c (\slashed{u} \Dslash \la )\\[0,1ex]\nonumber 
	{\cal O}_{C1} &=& {\rm{tr}}_c (\partial_{\mu}\la\, u^{\mu}) \\[0,1ex]\nonumber 
	{\cal O}_{C2} &=& {\rm{tr}}_c (\slashed{u} \la) \\[0,1ex]\nonumber 
	{\cal O}_{C3} &=& ig\,\sigma_{\mu \nu} {\rm{tr}}_c (\, \lambda [u_{\mu}, u_{\nu}] )\\[0,1ex]\nonumber 
	{\cal O}_{C4} &=& ig\, {\rm{tr}}_c(\la [c, \bar{c}])\nonumber 
	\label{All_operators}
	\eea
	Class C operators cannot contribute in the continuum for the purpose of $\MSbar$-renormalization. However, they may give finite mixing coefficients on the lattice. Note also that the operator ${\cal O}_{C2}$ is of lower dimension and it will not mix with ${\cal O}_{Gg}$ in dimensional regularization; it may however show up in the lattice formulation. The presence of symmetries, which are preserved by the SYM action, both in the continuum and on the lattice, forbids other operators from mixing with the Gluino-Glue operator. There are no gauge-invariant operators which mix with ${\cal O}_{Gg}$; thus, by applying Gauge Invariant Renormalization Scheme (GIRS)~\cite{Costa:2021iyv} we only need to calculate one Green's function (Eq.~(\ref{GFsGg})) to extract the multiplicative renormalization factor of ${\cal O}_{Gg}$, which contributes in physical matrix elements.
	
	\section{Renormalization of Gluino-Glue operator in the GIRS scheme}
\begin{figure}[]
		\centering
		\epsfig{file=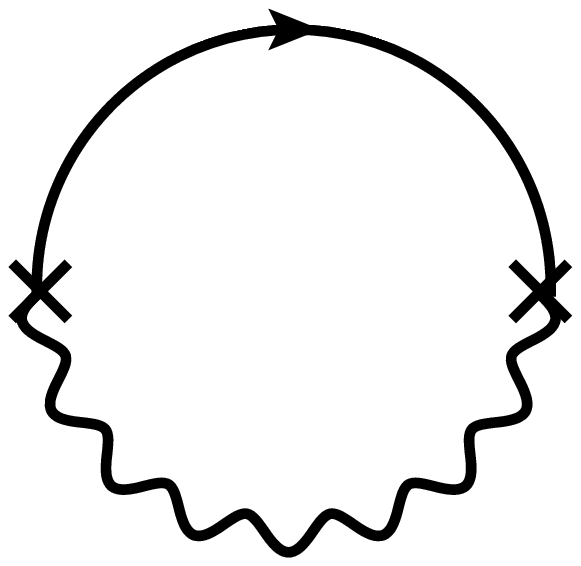,scale=0.211}\\
		\epsfig{file=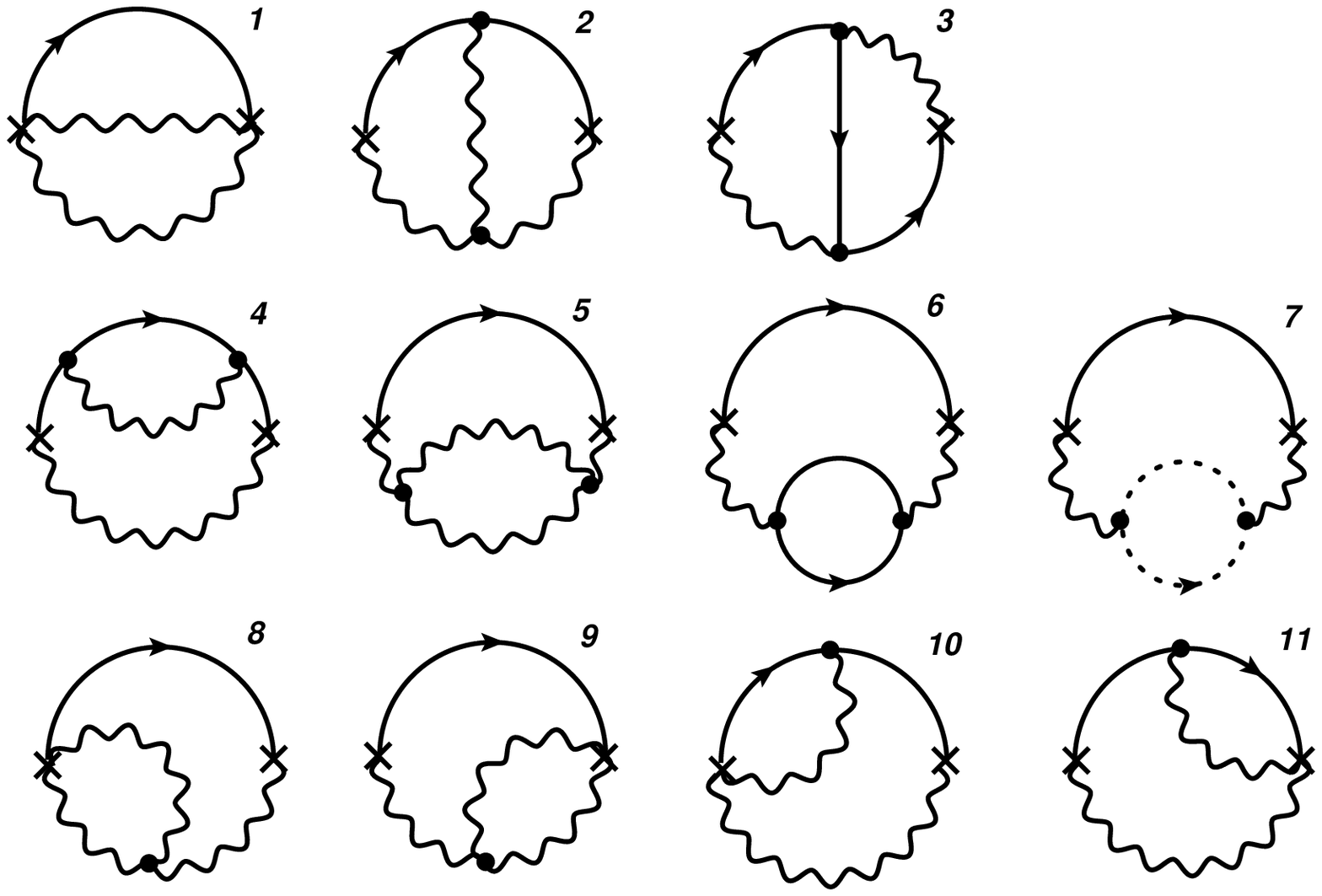,scale=0.334}
		\caption{Tree-level and (eleven) one-loop Feynman diagrams contributing to the expectation value $\langle {\cal O}_{Gg}(x) {\overline{\cal O}}_{Gg}(y)\rangle$. A wavy (solid) line represents gluons (gluinos). The dashed line is the ghost field. A cross denotes the insertion of the operator. }
		\label{treeANDoneloop}
	\end{figure}

	In the GIRS scheme, renormalization factors  are defined via the following Green's function, containing a product of Gluino-Glue operators, whose 4-vector positions $x$ and $y$ are distinct:
	\be
	G(x-y) \equiv \langle {\cal O}_{Gg} (x) {\overline{\cal O}}_{Gg} (y) \rangle. 
	\label{GFsGg}
	\ee
	In order to contract gluino fields in Feynman diagrams it is convenient to choose the charge conjugate operator ${\overline{\cal O}}_{Gg}(y)$, instead of ${\cal O}_{Gg}(y)$, as the second factor in Eq.~(\ref{GFsGg});  ${\overline{\cal O}}_{Gg}(y)$ is defined as (cf. Eq.~(\ref{GgO})):
	\be
	{\overline{\cal O}}_{Gg} =  -\,{\rm{tr}}_c (\, \bar \lambda u_{\mu \nu} ) \sigma_{\mu \nu}.
	\ee
	Disconnected Feynman diagrams are not present in $G(x-y)$ due to the fact that ${\cal O}_{Gg}$ is not a scalar operator. In the following, we calculate the Green's function up to one loop, where we regularize the theory in $d$ dimensions ($d=4-2\,\epsilon$). In the GIRS scheme all non-gauge invariant operators will not contribute to $G(x-y)$, and one will obtain directly the multiplicative renormalization of ${\cal O}_{Gg}$, which is the only renormalization factor which is relevant for physical matrix elements. The Feynman diagrams for the tree-level and one-loop values of the 2-pt Green's function are shown in Fig.~\ref{treeANDoneloop}. By adding tree-level and one-loop contributions, the bare Green's function takes the following form:
	\bea
	G(x-y)^{\rm bare}&& {=} 
	-2\, \frac{(N_c^2-1)\:\Gamma(2-\epsilon)^2}{\pi^{4-2\epsilon}} (-1 + \epsilon)(-3 + 2\epsilon) \,\slashed{z}\,(z^2)^{-4 + 2\epsilon} \times \nonumber \\
	&& \Bigg\{ 1 - \frac{g^2 \ N_c}{16 \pi^2} \left(\bar{\mu}^2 z^2 \right)^\epsilon \frac{e^{\epsilon \gamma_E} \Gamma(-\epsilon)} {4^\epsilon \ \epsilon \ (1 - \epsilon)^3 \ (3 - 2\epsilon)} \times  \nonumber \\
	&&\hspace{-2.25cm}  \Bigg[ (1-\epsilon) (12 -48 \epsilon + 70 \epsilon^2 - 39 \epsilon^3 + \epsilon^4) + \frac{(1 - 3 \epsilon + 2 \epsilon^2 + \epsilon^3) 
		\Gamma(-\epsilon) \Gamma(\epsilon)^2  \Gamma(4-3\epsilon)}{4  (1-2\epsilon)  \Gamma(-2\epsilon)^2 \Gamma(2\epsilon)} \Bigg] \Bigg\},
	\label{GFtreePoneloop}
	\eea
	where $z^\mu \equiv y^\mu - x^\mu$,  $\bar\mu$ is the $\MSbar$ renormalization scale relating the dimensionful coupling constant $g_L$ in the $d$-dimensional Lagrangian to the dimensionless ``bare'' coupling constant $g^B$: $g_L = \mu^\epsilon g^B$ ($\mu = \bar \mu \sqrt{e^{\gamma_E}/ 4\pi}$). To this perturbative order, the distinction between bare and renormalized coupling constants is inessential; we will thus denote both by $g$. 
	A more promising option is to integrate~Eq.(\ref{GFsGg}) over three of the four components of the position vector $(x - y)$, while setting the fourth component equal to a reference scale $\bar{t}$.  Due to the anisotropic lattice employed in simulations, the temporal direction is a special one. In this sense, a natural choice for the component $\bar{t}$ is to be temporal; we call this variant ``t-GIRS''. Without loss of generality, we set $x=(x_1,x_2,x_3,0)$ and $y=(0,0,0,\bar{t})$; then the renormalization condition for t-GIRS has the following form:
	\begin{equation}
	{\rm Tr} \Big[ \Big( \int d^3\vec{x} \ \langle {\mathcal O}^{\rm t-GIRS}_{Gg} (\vec{x}, 0) {{\overline{\mathcal O}}^{\rm t-GIRS}_{Gg}} (\vec{0},\bar{t}) \rangle \Big) \gamma_4 \Big] = (Z^{B,{\rm t-GIRS}}_{Gg})^2{\rm Tr} \Big[ \Big( \int d^3\vec{x} \ \langle {\mathcal O}^B_{Gg} (\vec{x}, 0) {{\overline{\mathcal O}}^B_{Gg}} (\vec{0}, \bar{t}) \rangle^{\rm tree} \Big) \gamma_4 \Big],
	\label{t-GIRS_condition}
	\end{equation}
	where the right-hand side of the above condition is the tree-level Green's function with $\epsilon \rightarrow 0$. The conversion factor between $\MSbar$ and t-GIRS, ${\mathcal O}^{\MSbar}_{Gg} = C_{Gg}^{ {\rm t-GIRS}, \MSbar} {\mathcal O}^{\rm t-GIRS}_{Gg}$, is obtained from the relation:
	\begin{equation}
	{\rm Tr} \Big[ \Big( \int d^3\vec{x} \ \langle {\mathcal O}^{\MSbar}_{Gg} (\vec{x}, 0) {{\overline{\mathcal O}}^{\MSbar}_{Gg}} (\vec{0},\bar{t}) \rangle \Big) \gamma_4 \Big] = \left(C_{Gg}^{\MSbar,{\rm t-GIRS}}\right)^2 {\rm Tr} \Big[ \Big( \int d^3\vec{x} \ \langle {\mathcal O}^{\rm t-GIRS}_{Gg} (\vec{x}, 0) {{\overline{\mathcal O}}^{\rm t-GIRS}_{Gg}} (\vec{0}, \bar{t}) \rangle^{\rm tree} \Big) \gamma_4 \Big].
	\label{t-GIRS_condition2}
	\end{equation}
	Note that both sides of Eq.~(\ref{t-GIRS_condition2}) are well-defined as $d \to 4$, and thus spatial integration is performed in $3$ (rather than $d-1$) dimensions. Use of the integrals:
	\be
	\int d^3\vec{x} \ \frac{1}{(|\vec{x}|^2 + t^2)^4} = \frac{\pi^2}{8 |t|^5}, \quad \int d^3\vec{x} \ \frac{\ln(|\vec{x}|^2 + t^2)}{(|\vec{x}|^2 + t^2)^4} = \frac{\pi^2 (-5 + 12 \ln(2) + 6 \ln (t^2))}{48 |t|^5},
	\ee
	in Eq.~(\ref{t-GIRS_condition2}), leads to the following expression for the conversion factor from  t-GIRS to $\MSbar$:
	\begin{equation}
	C_{Gg}^{\MSbar, {\rm t-GIRS}} = 1 + \frac{g^2 N_c}{16 \pi^2} \Big(-\frac{5}{6} + 6 \gamma_E + 3 \ln (\bar{\mu}^2 \bar{t}^2) \Big) + \mathcal{O} (g^4).
	\end{equation}
	Using a standard lattice discretization ${\cal O}_{Gg}^L$ of the Gluino-Glue operator~\cite{Ali:2018dnd}, the non-perturbative value of its renormalization factor, $Z_{Gg}^{L, {\rm t-GIRS}}$, can be found via:
	\bea
	(Z_{Gg}^{L,{\rm t-GIRS}})^2 {\rm Tr} \Big[ \Big( \int d^3\vec{x} \ \langle {\mathcal O}^{L}_{Gg} (\vec{x}, 0) {{\overline{\mathcal O}}^{L}_{Gg}} (\vec{0},\bar{t}) \rangle \Big) \gamma_4 \Big] &=&
	{\rm Tr} \Big[ \Big( \int d^3\vec{x} \, \lim_{\epsilon \to 0}\langle {\cal O}_{Gg} (x) {\overline{\cal O}}_{Gg} (y) \rangle^{{\rm tree}}  \Big) \gamma_4 \Big]  \nonumber\\ 
	&=& 3 \frac{(N_c^2-1)|t|}{\pi^{2} t^{5}}.
	\label{Gtree4dim}
	\eea
	
	\section{Perturbative study of the Gluino-Glue operator on the lattice}
	
	Given that ${\cal O}_{Gg}$ has a relatively high dimensionality, 7/2, there is a number of other (non-gauge invariant) composite operators with which ${\cal O}_{Gg}$ can, and will, mix~\cite{Costa:2020keq}; this becomes apparent when one calculates Green's functions with elementary external fields, as is done in a typical renormalization procedure. A proper treatment of this mixing entails studying the 2-pt and 3-pt Green's functions of ${\cal O}_{Gg}$ with external gluino and gluon fields. The renormalized fields and operators are related to bare ones through:
	\be
	u_{\mu}^R = \sqrt{Z_u}\,u^B_{\mu},\quad \la^R = \sqrt{Z_\la}\,\la^B,\quad c^R = \sqrt{Z_c}\,c^B, \quad
	{\cal O}^R_{Gg}  = Z_{Gg} {\cal O}^B_{Gg} + z_{A1}{\cal O}^B_{A1} + z_{B1} {\cal O}^B_{B1}  + \sum_{i=1}^{4} z_{Ci}{\cal O}^B_{Ci} 
	\ee
	We compute the one-loop quantum correction for the relevant Green's functions of the Gluino-Glue operator. This allows us to determine renormalization factors of the operator in the $\MSbar$ scheme, as well as the mixing coefficients for the other operators. To this end, our computations are performed using dimensional and lattice regularizations. We employ a standard discretization where gluinos are defined on lattice sites and gluons reside on the links of the lattice; the discretization is based on Wilson's formulation of non-supersymmetric gauge theories with clover improvement. Our perturbative results are analytic expressions depending on the number of colors, $N_c$, the clover coefficient, $c_{\rm SW}$, the Wilson parameter, $r= \pm 1$, the lattice spacing $a$, and the gauge parameter, $\alpha$. 
	
	To obtain $Z_{Gg}$ and all mixing coefficients, $z$, we have calculated, to one loop and in an arbitrary covariant gauge, the 2-pt (gluino-gluon) and 3-pt (gluino-gluon-gluon and gluino-ghost-antighost) bare amputated Green's functions of ${\cal O}_{Gg}$; these are related to the corresponding renormalized Green's functions through:
	\bea 
	\langle u_\nu^R \,{\cal O}^R_{Gg}\, \bar\lambda^R \rangle _{amp} &=& Z_\la^{-1/2} \,Z_u^{-1/2} Z_{Gg} \langle u_\nu^B \,{\cal O}^B_{Gg}\,\bar\lambda^B \rangle _{amp}
	+ z_{A1}  \langle u_\nu^B \, {\cal O}^B_{A1}\, \bar\lambda^B \rangle _{amp}^{tree} + z_{B1} \langle  u_\nu^B \, {\cal O}^B_{B1} \, \bar\lambda^B \rangle _{amp}^{tree}\nonumber\\ 
	&&+ z_{C1} \langle u_\nu^B \,{\cal O}^B_{C1} \, \bar\lambda^B \rangle _{amp}^{tree}+ z_{C2}\langle u_\nu^B  \, {\cal O}^B_{C2}\, \bar\lambda^B \rangle _{amp}^{tree} 
	\label{2ptGFexpr}
	\\
	\langle u_\nu^R\,u_\mu^R \,{\cal O}^R_{Gg}\, \bar\lambda^R \rangle _{amp} &=& Z_\la^{-1/2} \,Z_u^{-1} Z_{Gg} \langle u_\nu^B u_\mu^B \,{\cal O}^B_{Gg}\, \bar\lambda^B  \rangle _{amp} +  z_{B1}\langle u_\nu^B u_\mu^B \, {\cal O}^B_{B1}\, \bar\lambda^B \rangle _{amp}^{tree} \nonumber\\ 
	&&+ z_{C3}\langle u_\nu^B u_\mu^B \, {\cal O}^B_{C3}\, \bar\lambda^B \rangle _{amp}^{tree} 
	\label{3ptGFexprGGg}
	\\
	\langle c^R \,{\cal O}^R_{Gg}\,\bar c^R\, \bar\lambda^R \rangle _{amp} &=&  
	Z_c^{-1} \,Z_\la^{-1/2} Z_{Gg} \langle c^B \,{\cal O}^B_{Gg}\,\bar c^b \bar\lambda^B  \rangle _{amp} + z_{A1}\langle c^B \, {\cal O}^B_{A1}\,\bar c^B \bar\lambda^B \rangle _{amp}^{tree}\nonumber\\ 
	&& + z_{C4}\langle c^B \, {\cal O}^B_{C4}\,\bar c^B \bar\lambda^B \rangle _{amp}^{tree} 
	\label{3ptGFexpr2}
	\eea

\begin{figure}[ht!]
		\centering
		\epsfig{file=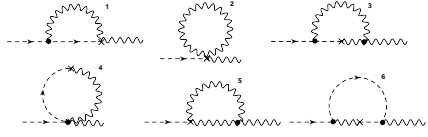,scale=0.334}
		\caption{One-loop Feynman diagrams contributing to the two point Green's
			function of the Gluino-Glue operator, $\langle u_\nu {\cal O}_{Gg} \bar \la  \rangle$\,.  A wavy (dashed) line represents gluons (gluinos). A cross denotes the insertion of the Gluino-Glue operator. Diagrams 2, 4 do not appear in dimensional regularization; they do however show up in the lattice formulation.}
		\label{fig2pt}
	\end{figure}

\begin{figure}[ht!]
		\centering
		\epsfig{file=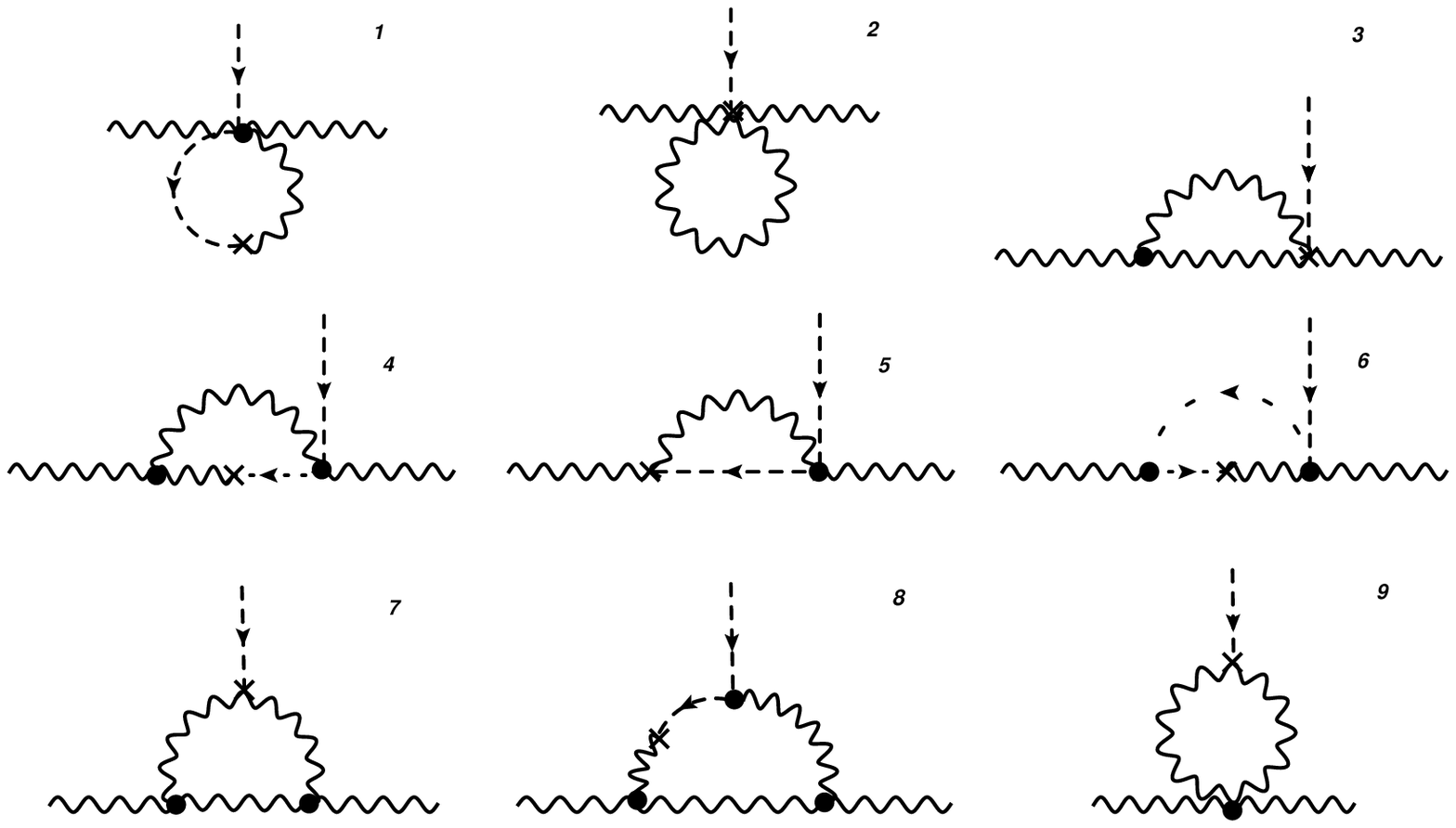,scale=0.334}\\
		\epsfig{file=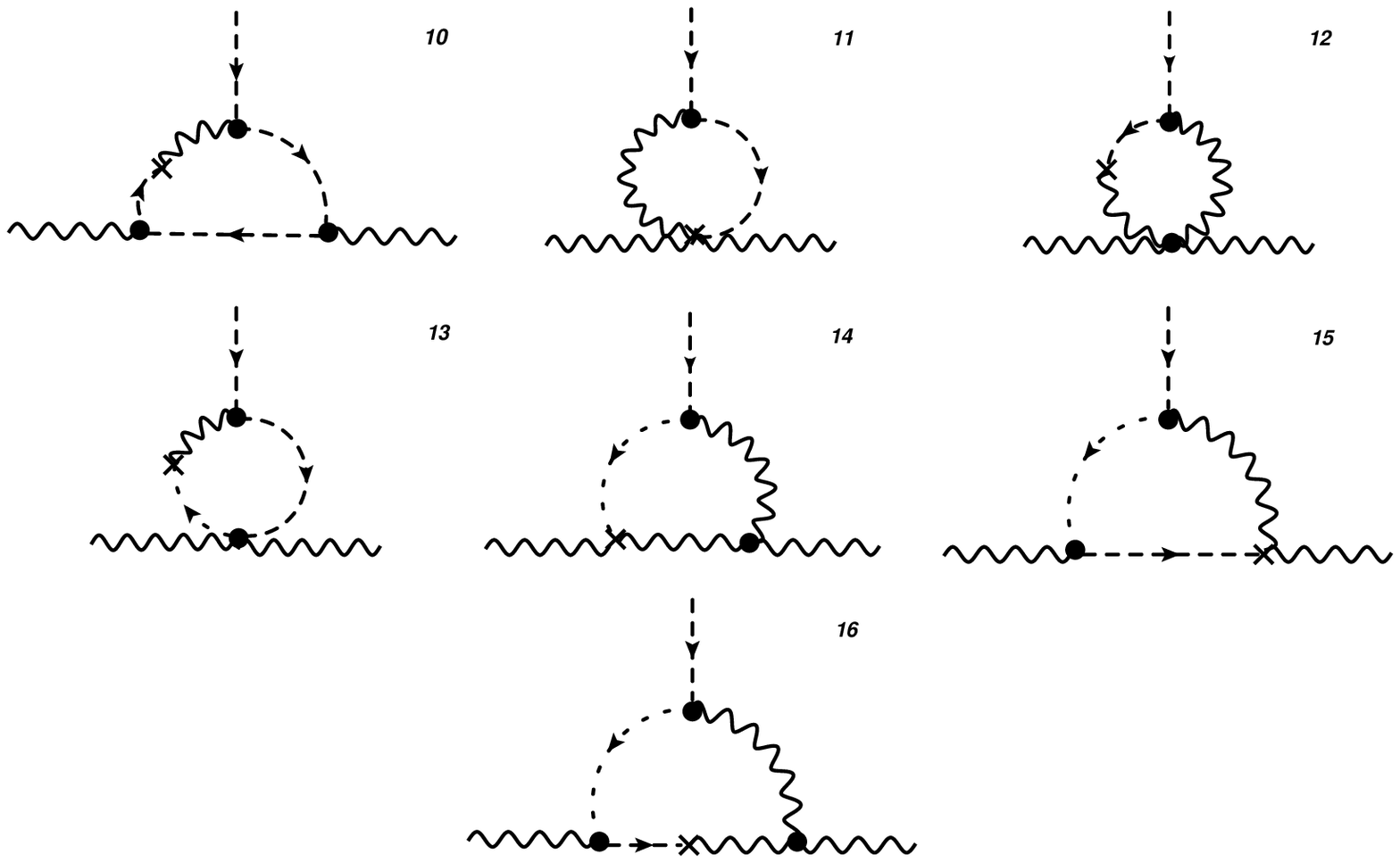,scale=0.334}
		\caption{One-loop Feynman diagrams contributing to the three point Green's
			function of the Gluino-Glue operator, $\langle u_\nu u_\mu {\cal O}_{Gg}\bar \la  \rangle$\,.  A wavy (dashed) line represents gluons (gluinos). Diagrams 1, 2, 3, 5, 6, 11, and 13 do not appear in dimensional regularization but they contribute in the lattice regularization. A cross denotes the insertion of the operator. A mirror version (under exchange of the two external gluons) of diagrams 3, 4, 5, 6, 8, 10, 14, 15 and 16 must also be included.}
		\label{fig3ptguu}
	\end{figure}

\begin{figure}[ht!]
		\centering
		\epsfig{file=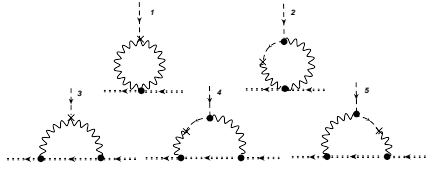,scale=0.334}
		\caption{One-loop Feynman diagrams contributing to the three point Green's
			function of the Gluino-glue operator, $\langle c \, {\cal O}_{Gg} \, \bar c \, \bar \la \rangle$\,.  A wavy (dashed) line represents gluons (gluinos). A cross denotes the insertion of the operator. The ``double dashed'' line is the ghost field. Diagrams 1 and 2 do not appear in dimensional regularization; they do however show up in the lattice formulation.}
		\label{fig3ptgCC}
	\end{figure}

	The one-loop Feynman diagrams (one-particle irreducible (1PI)) contributing to these Green's functions are shown in Figs.~\ref{fig2pt}, \ref{fig3ptguu}, \ref{fig3ptgCC}.
	We have made a choice of the external momenta for Green's functions which allow an unambiguous extraction of all mixing coefficients and renormalization constants of the operators. We calculate the 2-pt Green's function $\langle u_\nu^{\alpha_1}(-q_1) {\cal O}_{Gg}(x) \bar \la^{\alpha_2}(q_2)  \rangle$, for three choices of the external momenta $q_1$ and $q_2$: $q_2=0$, $q_1=0$ and $q_2=-q_1$. The 3-pt Green's functions $\langle c^{\alpha_3}(q_3) \,{\cal O}_{Gg}\,\bar c^{\alpha_2}(q_2) \bar\lambda^{\alpha_1}(q_1) \rangle$ and $\langle u_\nu^{\alpha_1}(-q_1) u_\mu^{\alpha_2}(-q_2) \,{\cal O}_{Gg}\, \bar\lambda^{\alpha_3}(q_3) \rangle$ are calculated at (${q_1=q_2,\, q_3 = 0}$) and (${q_2 =0 , q_3 = -q_1}$), respectively. We use the $\MSbar$ renormalization scheme, in which the left hand sides of the renormalization conditions in Eqs.~(\ref{2ptGFexpr}), (\ref{3ptGFexprGGg}) and (\ref{3ptGFexpr2}) are the $\MSbar$-renormalized Green's functions which are calculated by the elimination of the pole part of the continuum bare Green's functions. The right hand sides are the bare latice Green's functions. The relevant renormalizations of fields and gauge couplings constant were not all available for the clover action considered in this work, and had to be calculated as a prerequisite. Imposing the renormalization conditions we get the results for the renormalization factor and the mixing coefficients:
	\begin{eqnarray}
	Z_{Gg}^{L, \MSbar} &=& 1 - \frac{g^2 N_c}{16\pi^2}\bigg( \frac{9.8696}{N_c^2}-1.7626-9.9198\, c_{\rm SW}^2+4.9765\, c_{\rm SW}\, r   -3\log(a^2\,\bar\mu^2)\bigg)\\ 
	z_{B1}^{L, \overline{\textrm{MS}}} &=&  \frac{g^2 N_c}{16\pi^2}\bigg(0.4241 - \frac{3}{2}\log(a^2\,\bar\mu^2)\bigg)\\ 
	\label{ZB1L}
	z_{C3}^{L,\MSbar} &=&  - \frac{g^2 N_c}{16\pi^2} 0.000114
	\end{eqnarray}
	Systematic errors coming from numerical loop integration are much smaller than the precision presented in the above results. Also, certain mixing coefficients vanish at one loop:
	\be
	z_{A1}^{L,\MSbar} = z_{C1}^{L,\MSbar} = z_{C2}^{L,\MSbar}= z_{C4}^{L,\MSbar}=0
	\ee 
	
	Further details on this calculation can be found in Ref.~\cite{Costa:2020keq}.
	
	\section{Summary -- Future Plans}
	\label{summary}
	
	In this work we have studied the mixing under renormalization for the Gluino-Glue operator using lattice perturbation theory. We have calculated the one-loop renormalization factors and mixing coefficients in the $\MSbar$ renormalization scheme. As a prerequisite, we have computed the gluon, gluino and gauge coupling renormalization. Further, we have evaluated the conversion factor to $\MSbar$ for ${\cal O}_{Gg}$ to order $g^2$, using the GIRS renormalization scheme, by calculating the two-loop diagrams in the correlation function involving a product of two ${\cal O}_{Gg}$ operators at distinct positions. 
	
	Some natural extensions of our work, besides going to higher perturbative order, are to address SQCD, where the inclusion of quarks and squarks causes ${\cal O}_{Gg}$ to mix with other gauge invariant and noninvariant operators, even (on the lattice) ones of lower dimensionality. In our ongoing investigation we plan to address also other improved actions (e.g., overlap action, Wilson action using stout-smearing in the fermionic action) and other operators (e.g., Three-gluino operator, Noether Supercurrent operator). In particular, for detailed information on preliminary perturbative and non-perturbative results of the Noether supercurrent, see the proceedings by A. Skouroupathis and I. Soler Calero in this conference~\cite{Georg}.
	\\

	{\bf Acknowledgements:} M.C. and H.P. acknowledge financial support from the project ``Quantum Fields on the Lattice'', funded by the Cyprus Research and Innovation Foundation (RIF) under contract number EXCELLENCE/0918/0066.

\end{document}